\documentclass[sigconf]{acmart}
\AtBeginDocument{%
  }

\begin{document}

\title{Decoupled Entity Representation Learning  for \\ Pinterest Ads Ranking}

\author{Jie Liu}
\authornote{All three authors contributed equally to this research.}
\email{jieliu@pinterest.com}
\orcid{1234-5678-9012}
\affiliation{%
  \institution{Pinterest, Inc}
   \city{San Francisco}
  \country{USA}
}

\author{Yinrui Li}
\authornotemark[1]
\email{yinruili@pinterest.com}
\affiliation{%
  \institution{Pinterest, Inc}
   \city{San Francisco}
  \country{USA}
}

\author{Jiankai Sun}
\authornotemark[1]
\email{jiankaisun@pinterest.com}

\affiliation{%
  \institution{Pinterest, Inc} 
   \city{San Francisco}
  \country{USA}
}

\author{Kungang Li}
\email{kungangli@pinterest.com}
\affiliation{%
  \institution{Pinterest, Inc}
   \city{San Francisco}
  \country{USA}
}

\author{Han Sun}
\email{hsun@pinterest.com}
\affiliation{%
 \institution{Pinterest, Inc}
  \city{San Francisco}
 \country{USA}}

\author{Sihan Wang}
\email{sihanwang@pinterest.com}
\affiliation{%
 \institution{Pinterest, Inc}
  \city{San Francisco}
 \country{USA}}

\author{Huasen Wu}
\email{huasenwu@pinterest.com}
\affiliation{%
 \institution{Pinterest, Inc}
  \city{San Francisco}
 \country{USA}}

\author{Siyuan Gao}
\email{siyuangao@pinterest.com}
\affiliation{%
 \institution{Pinterest, Inc}
 \city{San Francisco}
 \country{USA}}
 
\author{Paulo Soares}
\email{pdasilvasoares@pinterest.com}
\affiliation{%
 \institution{Pinterest, Inc}
 \city{San Francisco}
 \country{USA}}

\author{Nan Li}
\email{nanli@pinterest.com}
\affiliation{%
 \institution{Pinterest, Inc}
 \city{San Francisco}
 \country{USA}}

\author{Zhifang Liu}
\email{zhifangliu@pinterest.com}
\affiliation{%
 \institution{Pinterest, Inc}
 \city{San Francisco}
 \country{USA}}

\author{Haoyang Li}
\email{haoyangli@pinterest.com}
\affiliation{%
 \institution{Pinterest, Inc}
 \city{San Francisco}
 \country{USA}}
 
\author{Siping Ji}
\email{siping@pinterest.com}
\affiliation{%
 \institution{Pinterest, Inc}
 \city{San Francisco}
 \country{USA}}

\author{Ling Leng}
\email{lleng@pinterest.com}
\affiliation{%
 \institution{Pinterest, Inc}
 \city{San Francisco}
 \country{USA}}

\author{Prathibha
Deshikachar}
\email{pdeshikachar@pinterest.com}
\affiliation{%
 \institution{Pinterest, Inc}
 \city{San Francisco}
 \country{USA}}

\renewcommand{\shortauthors}{Liu et al.}

\begin{abstract}
In this paper, we introduce a novel framework following an upstream-downstream paradigm 
to construct user and item (Pin) embeddings from diverse data sources, which are essential for 
Pinterest to deliver personalized Pins and ads effectively. 
Our upstream models are trained on extensive data sources featuring varied signals,
utilizing complex architectures to capture intricate relationships between users and 
Pins on Pinterest. To ensure scalability of the upstream models,
entity embeddings are learned, and regularly refreshed,
rather than real-time computation, allowing for asynchronous interaction between the upstream and downstream models.
These embeddings are then integrated as input features in numerous
downstream tasks, including ad retrieval and ranking models for CTR and CVR predictions. 
We demonstrate that our framework achieves notable performance improvements in 
both offline and online settings across various downstream tasks.
This framework has been deployed in Pinterest’s production ad ranking systems,
resulting in significant gains in online metrics.
\end{abstract}




\maketitle
\section{Introduction}
\label{sec:introduction}
Modern recommender systems depend extensively on different entity representations, such as those for users, items, and queries. These representations map entities into low-dimensional dense vector spaces for efficient processing by deep neural networks. In Pinterest's ads recommender system, these embeddings are extensively utilized throughout the ranking pipeline, serving key roles in candidate retrieval through Approximate Nearest Neighbor (ANN) searches \cite{johnson2017billionscalesimilaritysearchgpus}—and in ranking as important features.

Pinterest offers a variety of surfaces, such as Home Feed, RePin, and Search, to support various user-interaction scenarios. Users and advertisers also interact with ads in multiple optimization types, including clicks and conversions. Ideally, leveraging data across all these domains and ad types would yield richer, more effective entity representations. However, this is challenging: each surface and ad type comes with unique optimization objectives, separate datasets, and specialized models, resulting in fragmented representations and limited information sharing. Infrastructure limitations further exacerbate this by restricting models to a subset of features, limiting a holistic view of user behavior. Performing efficient knowledge transfer across these domains remains a significant hurdle. 

To address these challenges, we introduce a decoupled upstream-downstream framework that learns the embedding of user and Pin (item) from various data sources, critical for delivering personalized content and ads on Pinterest. Building on the approach outlined in \cite{zhuang2025}, our upstream models are trained on large-scale datasets with rich user interaction signals, utilizing advanced architectures to capture the deep relationships between users and Pins. For scalability, entity embeddings are pre-computed and regularly refreshed, rather than calculated in real time, which allows asynchronous serving between upstream and downstream models. These embeddings are then integrated as key features in a wide range of downstream models, including those that handle ad retrieval, CTR, and CVR predictions for ranking tasks. These embeddings have shown significant improvements in offline performance and have also been successfully deployed in production for CTR prediction. 

Compared with previous work \cite{Zhang_2024}, our work offers several key innovations.
\begin{itemize}
    \item We propose a new model architecture featuring multiple towers that produce distinct embeddings for entities such as users, Pins and others. Interaction features are provided to a specialized overtower, preventing contextual information from contaminating the core embeddings and thereby enhancing their stability.
    \item In addition to multi-task supervised learning, we leverage self-supervised learning on the outputs of the entity towers using a contrastive loss technique. 
    \item To incorporate smooth and performing embedding shift, we use a weighted moving average to combine current and historical embeddings. We find that assigning greater weight to past embeddings delivers the best performance improvement among other aggregation approaches in our setup. 
\end{itemize}

\section{Related Work}

Unlike the classic two-tower model that trains on similarity measures (e.g., dot product) of user and item embeddings \cite{Huang_2020}, our model employs a complex overtower to process both tower outputs and interaction features. A similar overtower design is present in \cite{Zhang_2024}, although that work only focuses on user representation learning. Beyond tower architectures, technologies like graph-based algorithms \cite{Ying_2018, hamilton2018inductiverepresentationlearninglarge} and noise contrastive loss \cite{El_Kishky_2022} enhance representation learning by exploring item-item relationships. Recently, transformers have been widely used to generate embeddings from user sequences \cite{Agarwal_2024, baltescu2022itemsagelearningproductembeddings}. Additionally, users can be characterized through the aggregation of item embeddings \cite{Pal_2020}. In contrast to methods that concentrate on a single technology, our work aims to develop a versatile framework that integrates diverse and rich signals, utilizing various supervised and unsupervised loss functions for training.

\section{Our Methods}
\label{sec:methods}
In this section, we present our methods for developing \textbf{D}ecoupled \textbf{E}ntity \textbf{R}epresentation \textbf{M}odel (\textbf{DERM}), and the utilization of DERM embeddings in downstream tasks.

\subsection{Upstream Model Architecture}
\label{sec:upstream_model}
\begin{figure}
\includegraphics[width=\columnwidth]{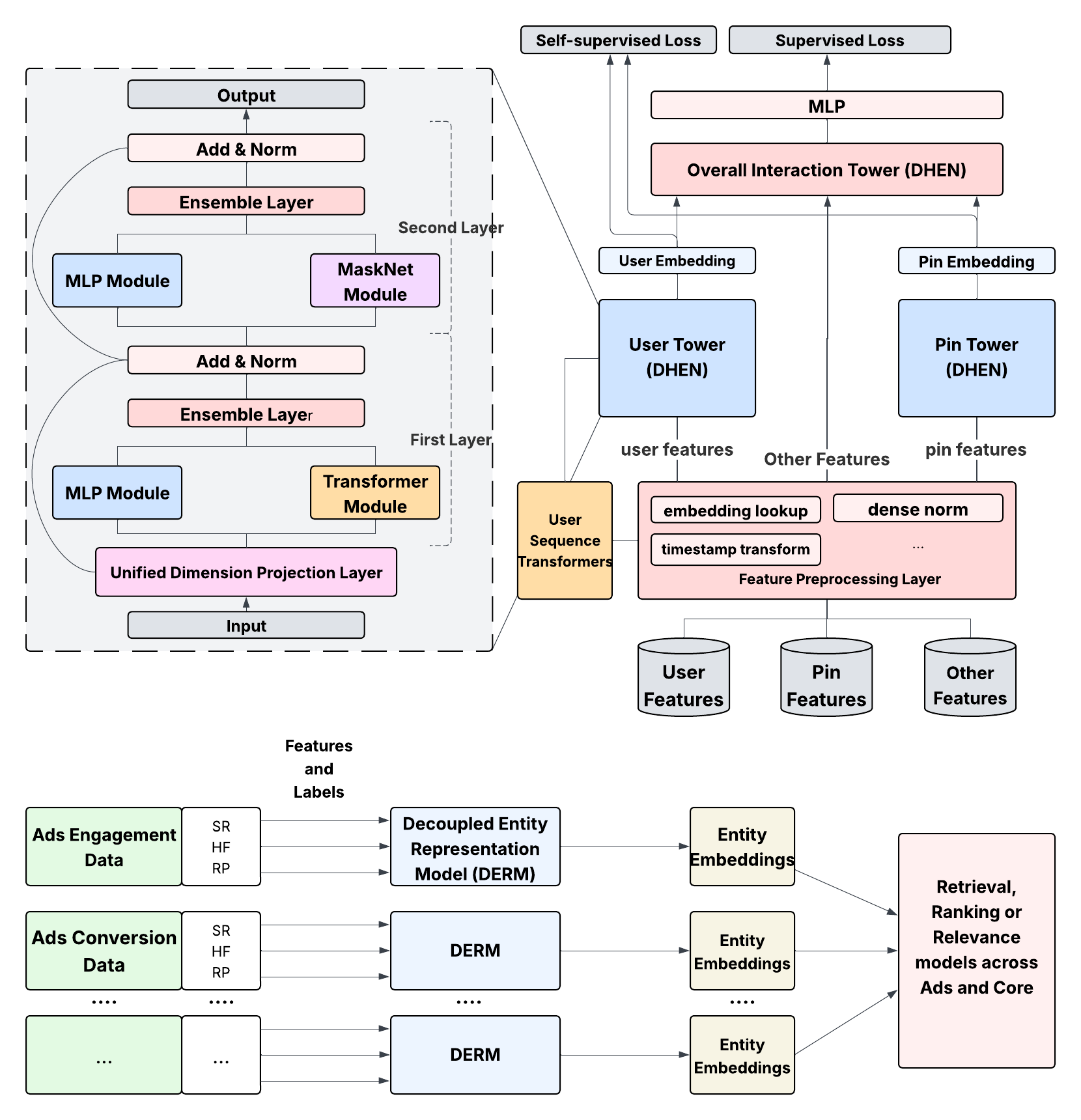}
\caption{(top) DERM model architecture and (bottom) upstream-downstream framework. The upstream model adopts a three-tower multi-task learning architecture. The user and Pin towers utilize a multi-layer DHEN backbone and processes the corresponding entity (user or ad Pin) features. Tower outputs along with other features are fed into the overall interaction tower for final crossing. To enhance model performance, we add a self-supervised learning task between the outputs of the two towers. The pre-trained embeddings are consumed by various downstream domains}
\label{fig:derm}
\end{figure}

DERM adopts a multi-tower architecture to generate embeddings for various entities (Figure~\ref{fig:derm}). Each tower uses a multi-layer Deep Hierarchical Ensemble Network (DHEN) backbone \cite{zhang2022dhendeephierarchicalensemble, zhuang2025}, with each layer consisting of two feature interaction modules chosen from Masknet \cite{wang2021masknetintroducingfeaturewisemultiplication}, Transformer\cite{vaswani2023attentionneed} and MLP.

We primarily utilize Pinterest's two extensive ad production datasets (CTR and CVR prediction) as data sources to create DERM embeddings. For each dataset, we develop a specific upstream model. Each model comprises three towers: 1) \textbf{User Tower}, which is crafted to process a diverse array of user features, including user activity sequences, demographics, curated interests, counting features, and pretrained representations. 2) \textbf{Pin Tower}, which ingests all features related to ad Pins, such as content metadata, aggregated counting features, pretrained representations, and additional relevant insights into the ad content. 3)  \textbf{Overall Interaction Tower}, which processes the interactions between the outputs of the above two towers along with additional contextual and interaction features.

To improve the learning of entity embeddings and the effectiveness of the upstream models, we add an additional self-supervised learning task between the user and ad Pin towers. When a training sample is linked to a positive click or conversion label, we consider the associated user and target Pin pair $(u_i, p_i)$ as positive. Drawing inspiration from previous research on pre-trained embeddings ~\cite{PanchaZLR22}\cite{radford2021learningtransferablevisualmodels}, for each positive user and target Pin pair, we generate a set of randomly sampled in-batch negative examples and employ the following sampled softmax loss 

\[
L_{u_i, p_i} =  -\log\left(\frac{e^{\mathbf{u}_{i}^\top \mathbf{p}_{i}/\tau - \log(Q(p_i))}}{e^{\mathbf{u}_{i}^\top \mathbf{p}_{i}/\tau - \log(Q(p_i))} + \sum_{p_{j} \in N_{i}^{-}}e^{ \mathbf{u}_{i}^\top \mathbf{p}_{j}/\tau- \log(Q(p_j))}}\right),
\]
where $\mathbf{u}_{i}$ denotes the user embedding, $\mathbf{p}_{i}$ denotes the target Pin embedding, $N_{i}^{-}$ denotes a set of randomly sampled negative Pins from a batch per user and target Pin pair $(u_i, p_i)$, and $\tau$ is a trainable temperature. We apply a sampling bias correction term $Q$ to each logit as an estimate of the frequency of the corresponding Pin ($p_i$ or $p_j$) is seen in a batch. The final loss is a combination of this sampled softmax loss with the losses of supervised learning tasks such as CVR or CTR prediction.

The DERM embeddings produced by the User and Pin towers serve as user and Pin embeddings for downstream applications, respectively. These entity embeddings are generated daily from their respective data sources in an offline process. To maintain coverage and stability, we apply a moving average technique that computes a weighted sum of the previous day's embedding and the newly generated embedding. This is represented by $E_{\mathrm{agg}}(t) =wE_{\mathrm{agg}}(t-1) + (1 - w)E_{\mathrm{daily}}(t)$ where $t$ denotes the $t$-th day and the weight $w$ is tuned as $0.8$. The aggregated embeddings are subsequently uploaded to a key-value feature store for online serving and downstream tasks.
 
\subsection{Downstream Tasks}
\label{sec:downstream_model}

In this section, we will explore the utilization of the DERM embeddings
generated by the upstream models for Pinterest's downstream tasks. Specifically,
we have two downstream tasks: (1) CTR predictions and (2) CVR predictions.

Our CTR model is built as a multitask learning framework that integrates multiple
objectives within a single unified model. Each task is linked to its own tower
to make specific predictions. The core of the model is a Mixture of Experts
(MoE)~\cite{mmoe1991,mmoe1993} module, which utilizes DCNv2 ~\cite{DCNv22021} to perform feature interactions. Prior to the
MoE, there is a concatenation layer that merges embeddings from
various sources, such as users' categorical and continuous features. A transformer encoder ~\cite{zhuang2025} handles user sequential information, generating embeddings
that are then concatenated with others to be placed in the concatenation layer.
Additionally, DERM embeddings are combined with existing embeddings before being
fed into the concatenation layer. 
Our CVR model has a similar architecture as the upstream model based on DHEN, as detailed in ~\cite{zhuang2025}\cite{qiu2025evolutionembeddingtableoptimization}. We employ the same strategy as described above for utilizing the DERM embeddings in the CVR model.

\section{Experimental Results}
\label{sec:experiments}

In this section, we present the results of our experiments, which were conducted
in both offline and online settings.

\subsection{Results in Offline Experiments}
\label{sec:offline-evaluation}
\subsubsection{Aggregation Heuristics}
\label{sec:aggregation_heuristics_experiment}

We experimented with various aggregation heuristics: 1) Accumulation (ACC), which retains the most recent embedding. 2) Moving Average (MA), as outlined in Section ~\ref{sec:upstream_model}, with varying  $w$. 3) Average Pooling (AP), which averages the two most recent daily embeddings. We illustrate the impact of these strategies by applying them to CVR user embeddings and evaluating the results on downstream CVR tasks. As depicted in Table \ref{tab:aggregation_schemes_auc}, we can see that the MA with $w=0.8$ performs the best. Similar trends can be observed with other DERM embeddings in different downstream tasks.

\begin{table}[h]
\centering
\captionsetup[table]{skip=1pt}
\caption{\small Offline AUC Lift from Aggregation Heuristics}
\label{tab:aggregation_schemes_auc}
\small
\begin{tabular}{l l l l l l}
\toprule
 ACC & MA $w=0.2$ & MA $w=0.5$ & MA $w=0.8$ & AP \\
\midrule
 0.02\% & 0.03\% & 0.06\% & \textbf{0.09}\% & 0.03\% \\
\bottomrule
\end{tabular}
\end{table}

\subsubsection{Number of DERM Embeddings}

In this section, we conducted experiments to evaluate the impact of different
DERM embeddings on the CTR task by testing four types of embeddings: user and item
embeddings from the upstream CTR model, and user and item embeddings from the
upstream CVR model. The results, presented in Table
\ref{tab:number_of_derm_embeddings_ctr}, demonstrate that the combination of user
and item embeddings from both models yields the best performance on Pinterest's Home Feed
surface \footnote{\url{https://www.pinterest.com/}}. We noticed a similar pattern with our CVR tasks. The corresponding results are included in Appendix ~\ref{sec:number_of_derm_cvr} due to space constraints.

\begin{table}[h]\centering
  \caption{Sensitivity of Number of DERM Embeddings on Downstream Task (CTR Prediction) evaluated by ROC-AUC and PR-AUC Lift. A 
  checkmark (\checkmark) indicates that the corresponding embedding is used in the model.
  }
  \label{tab:number_of_derm_embeddings_ctr}
  \scriptsize
  \begin{tabular}{c|c|c|c||c|c}\toprule
    \multicolumn{4}{c||}{Upstream Models} & \multicolumn{2}{c}{Downstream Task: CTR AUC } \\
    \midrule
    \multicolumn{2}{c|}{CTR} & \multicolumn{2}{c||}{CVR} & ROC-AUC Lift (\%) & PR-AUC Lift (\%) \\ \hline
    User & Item & User & Item &  \\ \hline
    \checkmark & \textbf{X} & \textbf{X} & \textbf{X} & 0.095 & 0.2 \\ \hline
    \textbf{X} & \textbf{X} & \checkmark & \textbf{X} & 0.059 & 0.14 \\ \hline
    \checkmark & \checkmark & \textbf{X} & \textbf{X} & 0.160 & 0.4 \\ \hline
    \checkmark & \textbf{X} & \checkmark & \textbf{X} & 0.120 & 0.25 \\ \hline
    \textbf{X} & \textbf{X} & \checkmark & \checkmark & 0.083 & 0.19 \\ \hline
    \checkmark & \checkmark & \checkmark & \checkmark & 0.190 & 0.46 \\ \hline
    \bottomrule
  \end{tabular}
\end{table}

\subsection{Results in Online Experiments}
\label{sec:online-evaluation}

We conducted online A/B tests to evaluate the impact of DERM embeddings
on the downstream CTR and CVR prediction tasks. Each experiment included a control group representing the production model and a treatment group using the DERM embeddings. 

\subsubsection{CTR Prediction}

Each group of CTR prediction accounted for $10\%$ of the total traffic on Pinterest's Related Pins Surface.
As shown in Table \ref{tab:online_ab_test_ctr}, we observe
that the treatment group achieved significantly larger CTR and gCTR\footnote{gCTR is good click through rate = $\#$ good clicks (i.e. clicks with an event duration of larger than 30 seconds / $\#$ impressions} lift compared to the control group. 

\begin{table}[!htp]
    \centering
    \caption{Online CTR and gCTR lift (all stats-sig)}
    \label{tab:online_ab_test_ctr}
    \scriptsize
    \renewcommand{\arraystretch}{1.25} 
\begin{tabular}{@{}lcccc@{}}
\toprule
            & \textbf{Platform-wise} & \textbf{RP Surface Only} & \textbf{Shopping Ads} & \textbf{Non-Shopping Ads} \\
\midrule
\textbf{CTR (\%) }   & 1.38 & 2.83 & 1.31 & 1.14 \\
\textbf{gCTR (\%)}  & 1.96 & 3.74 & 1.83 & 1.83 \\
\bottomrule
\end{tabular}
\end{table}
\subsubsection{CVR Prediction}

Similarly, we conducted online A/B testing experiments to evaluate the impact of DERM embeddings
on the downstream CVR task. Each group accounted for $9\%$ of the total traffic on Pinterest's traffic for web conversion optimization ads campaign. On all surfaces, we see $\mathbf{-1.61}\%$ CPA \footnote{total ads spend divided by total conversions driven by the conversion model. The less the better.} reduction. The click conversion rate (cCVR) \footnote{total conversions attributed to ads click divided by number of clicks} increased by $\mathbf{2.75}\%$ and the view conversion rate (vCVR) \footnote{total conversions attributed to ads viewing divided by number of impressions} increased by $\mathbf{2.8}\%$.

\section{Conclusion and Future Work}
\label{sec_conclusion}

In this paper, we have proposed a framework that enables the learning of DERM
embeddings (including both user and item) from multiple data sources. Extensive experiments were conducted on Pinterest's data, both
offline and online, to demonstrate the effectiveness of the learned DERM embeddings.
Moving forward, our future work will focus on updating the upstream model to incorporate
additional data sources and improve the efficiency of DERM embedding learning,
such as through the use of a single upstream model to learn the DERM embeddings
in a unified space. Additionally, we aim to develop content-based embeddings by feeding only non-ID features into the entity towers, thereby further enhancing the capability for cross-domain transfer learning.

\section*{SPEAKER BIO}

\textbf{Jie Liu} is currently a machine learning engineer at Pinterest where he works on using latest ML technologies to improve recommendation models. Previously, he worked as a data scientist at Oracle Inc, focusing on integrating machine learning technologies with autonomous data warehouse development cycles. He received Ph.D. from University of Notre Dame, with a focus on networked control and information theory. 

\noindent\textbf{Yinrui Li} is currently a machine learning engineer at Pinterest where she works on using latest ML technologies to improve recommendation models. Previously, she worked as a research professional at the University of Chicago Booth School of Business. She received her M.S. from University of Illinois Urbana-Champaign, with a research focus on simulations of particles dynamics in the atmosphere.

\noindent\textbf{Jiankai Sun} is currently a machine learning engineer at Pinterest where he works on using latest ML technologies to improve recommendation models. Previously, he worked as a research scientist at Bytedance Inc. Seattle, focusing on large language models and privacy-preserving techniques such as federated learning. He received Ph.D. from the Ohio State University with a focus on graph embedding and recommender systems.

\noindent\textbf{Kungang Li} is currently a senior technical leader at Pinterest SMLE (Science $\&$ ML Engineering), where he leads multiple ML areas for ads delivery. Prior to Pinterest, he has worked at Meta Ads Core ML and led data-centric AI. He received Ph.D. from Georgia Tech, with research focus on modeling and simulating particle dynamics in nano-scale.

\begin{acks}
We gratefully thank Youngjin Yun, Randy Carlson, Liangzhe Chen, Qifei Shen and Dongtao Liu's contributions to this work.
\end{acks}

\bibliographystyle{ACM-Reference-Format}
\bibliography{main}


\begin{thebibliography}{19}


\ifx \showCODEN    \undefined \def \showCODEN     #1{\unskip}     \fi
\ifx \showISBNx    \undefined \def \showISBNx     #1{\unskip}     \fi
\ifx \showISBNxiii \undefined \def \showISBNxiii  #1{\unskip}     \fi
\ifx \showISSN     \undefined \def \showISSN      #1{\unskip}     \fi
\ifx \showLCCN     \undefined \def \showLCCN      #1{\unskip}     \fi
\ifx \shownote     \undefined \def \shownote      #1{#1}          \fi
\ifx \showarticletitle \undefined \def \showarticletitle #1{#1}   \fi
\ifx \showURL      \undefined \def \showURL       {\relax}        \fi
\providecommand\bibfield[2]{#2}
\providecommand\bibinfo[2]{#2}
\providecommand\natexlab[1]{#1}
\providecommand\showeprint[2][]{arXiv:#2}

\bibitem[Agarwal et~al\mbox{.}(2024)]%
        {Agarwal_2024}
\bibfield{author}{\bibinfo{person}{Prabhat Agarwal},
  \bibinfo{person}{Minhazul~Islam SK}, \bibinfo{person}{Nikil Pancha},
  \bibinfo{person}{Kurchi~Subhra Hazra}, \bibinfo{person}{Jiajing Xu}, {and}
  \bibinfo{person}{Chuck Rosenberg}.} \bibinfo{year}{2024}\natexlab{}.
\newblock \showarticletitle{OmniSearchSage: Multi-Task Multi-Entity Embeddings
  for Pinterest Search}. In \bibinfo{booktitle}{\emph{Companion Proceedings of
  the ACM Web Conference 2024}} \emph{(\bibinfo{series}{WWW ’24})}.
  \bibinfo{publisher}{ACM}, \bibinfo{pages}{121–130}.
\newblock
\href{https://doi.org/10.1145/3589335.3648309}{doi:\nolinkurl{10.1145/3589335.3648309}}


\bibitem[Baltescu et~al\mbox{.}(2022)]%
        {baltescu2022itemsagelearningproductembeddings}
\bibfield{author}{\bibinfo{person}{Paul Baltescu}, \bibinfo{person}{Haoyu
  Chen}, \bibinfo{person}{Nikil Pancha}, \bibinfo{person}{Andrew Zhai},
  \bibinfo{person}{Jure Leskovec}, {and} \bibinfo{person}{Charles Rosenberg}.}
  \bibinfo{year}{2022}\natexlab{}.
\newblock \bibinfo{title}{ItemSage: Learning Product Embeddings for Shopping
  Recommendations at Pinterest}.
\newblock
\showeprint[arxiv]{2205.11728}~[cs.IR]
\urldef\tempurl%
\url{https://arxiv.org/abs/2205.11728}
\showURL{%
\tempurl}


\bibitem[El-Kishky et~al\mbox{.}(2022)]%
        {El_Kishky_2022}
\bibfield{author}{\bibinfo{person}{Ahmed El-Kishky}, \bibinfo{person}{Thomas
  Markovich}, \bibinfo{person}{Serim Park}, \bibinfo{person}{Chetan Verma},
  \bibinfo{person}{Baekjin Kim}, \bibinfo{person}{Ramy Eskander},
  \bibinfo{person}{Yury Malkov}, \bibinfo{person}{Frank Portman},
  \bibinfo{person}{Sofía Samaniego}, \bibinfo{person}{Ying Xiao}, {and}
  \bibinfo{person}{Aria Haghighi}.} \bibinfo{year}{2022}\natexlab{}.
\newblock \showarticletitle{TwHIN: Embedding the Twitter Heterogeneous
  Information Network for Personalized Recommendation}. In
  \bibinfo{booktitle}{\emph{Proceedings of the 28th ACM SIGKDD Conference on
  Knowledge Discovery and Data Mining}} \emph{(\bibinfo{series}{KDD ’22})}.
  \bibinfo{publisher}{ACM}, \bibinfo{pages}{2842–2850}.
\newblock
\href{https://doi.org/10.1145/3534678.3539080}{doi:\nolinkurl{10.1145/3534678.3539080}}


\bibitem[Hamilton et~al\mbox{.}(2018)]%
        {hamilton2018inductiverepresentationlearninglarge}
\bibfield{author}{\bibinfo{person}{William~L. Hamilton}, \bibinfo{person}{Rex
  Ying}, {and} \bibinfo{person}{Jure Leskovec}.}
  \bibinfo{year}{2018}\natexlab{}.
\newblock \bibinfo{title}{Inductive Representation Learning on Large Graphs}.
\newblock
\showeprint[arxiv]{1706.02216}~[cs.SI]
\urldef\tempurl%
\url{https://arxiv.org/abs/1706.02216}
\showURL{%
\tempurl}


\bibitem[Huang et~al\mbox{.}(2020)]%
        {Huang_2020}
\bibfield{author}{\bibinfo{person}{Jui-Ting Huang}, \bibinfo{person}{Ashish
  Sharma}, \bibinfo{person}{Shuying Sun}, \bibinfo{person}{Li Xia},
  \bibinfo{person}{David Zhang}, \bibinfo{person}{Philip Pronin},
  \bibinfo{person}{Janani Padmanabhan}, \bibinfo{person}{Giuseppe Ottaviano},
  {and} \bibinfo{person}{Linjun Yang}.} \bibinfo{year}{2020}\natexlab{}.
\newblock \showarticletitle{Embedding-based Retrieval in Facebook Search}. In
  \bibinfo{booktitle}{\emph{Proceedings of the 26th ACM SIGKDD International
  Conference on Knowledge Discovery \& Data Mining}}
  \emph{(\bibinfo{series}{KDD ’20})}. \bibinfo{publisher}{ACM}.
\newblock
\href{https://doi.org/10.1145/3394486.3403305}{doi:\nolinkurl{10.1145/3394486.3403305}}


\bibitem[Jacobs et~al\mbox{.}(1991)]%
        {mmoe1991}
\bibfield{author}{\bibinfo{person}{Robert~A. Jacobs},
  \bibinfo{person}{Michael~I. Jordan}, \bibinfo{person}{Steven~J. Nowlan},
  {and} \bibinfo{person}{Geoffrey~E. Hinton}.} \bibinfo{year}{1991}\natexlab{}.
\newblock \showarticletitle{Adaptive Mixtures of Local Experts}.
\newblock \bibinfo{journal}{\emph{Neural Computation}} \bibinfo{volume}{3},
  \bibinfo{number}{1} (\bibinfo{year}{1991}), \bibinfo{pages}{79--87}.
\newblock
\href{https://doi.org/10.1162/neco.1991.3.1.79}{doi:\nolinkurl{10.1162/neco.1991.3.1.79}}


\bibitem[Johnson et~al\mbox{.}(2017)]%
        {johnson2017billionscalesimilaritysearchgpus}
\bibfield{author}{\bibinfo{person}{Jeff Johnson}, \bibinfo{person}{Matthijs
  Douze}, {and} \bibinfo{person}{Hervé Jégou}.}
  \bibinfo{year}{2017}\natexlab{}.
\newblock \bibinfo{title}{Billion-scale similarity search with GPUs}.
\newblock
\showeprint[arxiv]{1702.08734}~[cs.CV]
\urldef\tempurl%
\url{https://arxiv.org/abs/1702.08734}
\showURL{%
\tempurl}


\bibitem[Jordan and Jacobs(1993)]%
        {mmoe1993}
\bibfield{author}{\bibinfo{person}{M.I. Jordan} {and} \bibinfo{person}{R.A.
  Jacobs}.} \bibinfo{year}{1993}\natexlab{}.
\newblock \showarticletitle{Hierarchical mixtures of experts and the EM
  algorithm}. In \bibinfo{booktitle}{\emph{Proceedings of 1993 International
  Conference on Neural Networks (IJCNN-93-Nagoya, Japan)}},
  Vol.~\bibinfo{volume}{2}. \bibinfo{pages}{1339--1344 vol.2}.
\newblock
\href{https://doi.org/10.1109/IJCNN.1993.716791}{doi:\nolinkurl{10.1109/IJCNN.1993.716791}}


\bibitem[Pal et~al\mbox{.}(2020)]%
        {Pal_2020}
\bibfield{author}{\bibinfo{person}{Aditya Pal}, \bibinfo{person}{Chantat
  Eksombatchai}, \bibinfo{person}{Yitong Zhou}, \bibinfo{person}{Bo Zhao},
  \bibinfo{person}{Charles Rosenberg}, {and} \bibinfo{person}{Jure Leskovec}.}
  \bibinfo{year}{2020}\natexlab{}.
\newblock \showarticletitle{PinnerSage: Multi-Modal User Embedding Framework
  for Recommendations at Pinterest}. In \bibinfo{booktitle}{\emph{Proceedings
  of the 26th ACM SIGKDD International Conference on Knowledge Discovery \&
  Data Mining}} \emph{(\bibinfo{series}{KDD ’20})}. \bibinfo{publisher}{ACM}.
\newblock
\href{https://doi.org/10.1145/3394486.3403280}{doi:\nolinkurl{10.1145/3394486.3403280}}


\bibitem[Pancha et~al\mbox{.}(2022)]%
        {PanchaZLR22}
\bibfield{author}{\bibinfo{person}{Nikil Pancha}, \bibinfo{person}{Andrew
  Zhai}, \bibinfo{person}{Jure Leskovec}, {and} \bibinfo{person}{Charles
  Rosenberg}.} \bibinfo{year}{2022}\natexlab{}.
\newblock \showarticletitle{PinnerFormer: Sequence Modeling for User
  Representation at Pinterest}. In \bibinfo{booktitle}{\emph{KDD}}.
  \bibinfo{publisher}{{ACM}}, \bibinfo{pages}{3702--3712}.
\newblock


\bibitem[Qiu et~al\mbox{.}(2025)]%
        {qiu2025evolutionembeddingtableoptimization}
\bibfield{author}{\bibinfo{person}{Andrew Qiu}, \bibinfo{person}{Shubham
  Barhate}, \bibinfo{person}{Hin~Wai Lui}, \bibinfo{person}{Runze Su},
  \bibinfo{person}{Rafael~Rios Müller}, \bibinfo{person}{Kungang Li},
  \bibinfo{person}{Ling Leng}, \bibinfo{person}{Han Sun},
  \bibinfo{person}{Shayan Ehsani}, {and} \bibinfo{person}{Zhifang Liu}.}
  \bibinfo{year}{2025}\natexlab{}.
\newblock \bibinfo{title}{The Evolution of Embedding Table Optimization and
  Multi-Epoch Training in Pinterest Ads Conversion}.
\newblock
\showeprint[arxiv]{2505.05605}~[cs.LG]
\urldef\tempurl%
\url{https://arxiv.org/abs/2505.05605}
\showURL{%
\tempurl}


\bibitem[Radford et~al\mbox{.}(2021)]%
        {radford2021learningtransferablevisualmodels}
\bibfield{author}{\bibinfo{person}{Alec Radford}, \bibinfo{person}{Jong~Wook
  Kim}, \bibinfo{person}{Chris Hallacy}, \bibinfo{person}{Aditya Ramesh},
  \bibinfo{person}{Gabriel Goh}, \bibinfo{person}{Sandhini Agarwal},
  \bibinfo{person}{Girish Sastry}, \bibinfo{person}{Amanda Askell},
  \bibinfo{person}{Pamela Mishkin}, \bibinfo{person}{Jack Clark},
  \bibinfo{person}{Gretchen Krueger}, {and} \bibinfo{person}{Ilya Sutskever}.}
  \bibinfo{year}{2021}\natexlab{}.
\newblock \bibinfo{title}{Learning Transferable Visual Models From Natural
  Language Supervision}.
\newblock
\showeprint[arxiv]{2103.00020}~[cs.CV]
\urldef\tempurl%
\url{https://arxiv.org/abs/2103.00020}
\showURL{%
\tempurl}


\bibitem[Vaswani et~al\mbox{.}(2023)]%
        {vaswani2023attentionneed}
\bibfield{author}{\bibinfo{person}{Ashish Vaswani}, \bibinfo{person}{Noam
  Shazeer}, \bibinfo{person}{Niki Parmar}, \bibinfo{person}{Jakob Uszkoreit},
  \bibinfo{person}{Llion Jones}, \bibinfo{person}{Aidan~N. Gomez},
  \bibinfo{person}{Lukasz Kaiser}, {and} \bibinfo{person}{Illia Polosukhin}.}
  \bibinfo{year}{2023}\natexlab{}.
\newblock \bibinfo{title}{Attention Is All You Need}.
\newblock
\showeprint[arxiv]{1706.03762}~[cs.CL]
\urldef\tempurl%
\url{https://arxiv.org/abs/1706.03762}
\showURL{%
\tempurl}


\bibitem[Wang et~al\mbox{.}(2021b)]%
        {DCNv22021}
\bibfield{author}{\bibinfo{person}{Ruoxi Wang}, \bibinfo{person}{Rakesh
  Shivanna}, \bibinfo{person}{Derek Cheng}, \bibinfo{person}{Sagar Jain},
  \bibinfo{person}{Dong Lin}, \bibinfo{person}{Lichan Hong}, {and}
  \bibinfo{person}{Ed Chi}.} \bibinfo{year}{2021}\natexlab{b}.
\newblock \showarticletitle{DCN V2: Improved Deep \& Cross Network and
  Practical Lessons for Web-scale Learning to Rank Systems}. In
  \bibinfo{booktitle}{\emph{Proceedings of the Web Conference 2021}}
  (Ljubljana, Slovenia) \emph{(\bibinfo{series}{WWW '21})}.
  \bibinfo{publisher}{Association for Computing Machinery},
  \bibinfo{address}{New York, NY, USA}, \bibinfo{pages}{1785–1797}.
\newblock
\showISBNx{9781450383127}
\href{https://doi.org/10.1145/3442381.3450078}{doi:\nolinkurl{10.1145/3442381.3450078}}


\bibitem[Wang et~al\mbox{.}(2021a)]%
        {wang2021masknetintroducingfeaturewisemultiplication}
\bibfield{author}{\bibinfo{person}{Zhiqiang Wang}, \bibinfo{person}{Qingyun
  She}, {and} \bibinfo{person}{Junlin Zhang}.}
  \bibinfo{year}{2021}\natexlab{a}.
\newblock \bibinfo{title}{MaskNet: Introducing Feature-Wise Multiplication to
  CTR Ranking Models by Instance-Guided Mask}.
\newblock
\showeprint[arxiv]{2102.07619}~[cs.IR]
\urldef\tempurl%
\url{https://arxiv.org/abs/2102.07619}
\showURL{%
\tempurl}


\bibitem[Ying et~al\mbox{.}(2018)]%
        {Ying_2018}
\bibfield{author}{\bibinfo{person}{Rex Ying}, \bibinfo{person}{Ruining He},
  \bibinfo{person}{Kaifeng Chen}, \bibinfo{person}{Pong Eksombatchai},
  \bibinfo{person}{William~L. Hamilton}, {and} \bibinfo{person}{Jure
  Leskovec}.} \bibinfo{year}{2018}\natexlab{}.
\newblock \showarticletitle{Graph Convolutional Neural Networks for Web-Scale
  Recommender Systems}. In \bibinfo{booktitle}{\emph{Proceedings of the 24th
  ACM SIGKDD International Conference on Knowledge Discovery \& Data Mining}}
  \emph{(\bibinfo{series}{KDD ’18})}. \bibinfo{publisher}{ACM},
  \bibinfo{pages}{974–983}.
\newblock
\href{https://doi.org/10.1145/3219819.3219890}{doi:\nolinkurl{10.1145/3219819.3219890}}


\bibitem[Zhang et~al\mbox{.}(2022)]%
        {zhang2022dhendeephierarchicalensemble}
\bibfield{author}{\bibinfo{person}{Buyun Zhang}, \bibinfo{person}{Liang Luo},
  \bibinfo{person}{Xi Liu}, \bibinfo{person}{Jay Li}, \bibinfo{person}{Zeliang
  Chen}, \bibinfo{person}{Weilin Zhang}, \bibinfo{person}{Xiaohan Wei},
  \bibinfo{person}{Yuchen Hao}, \bibinfo{person}{Michael Tsang},
  \bibinfo{person}{Wenjun Wang}, \bibinfo{person}{Yang Liu},
  \bibinfo{person}{Huayu Li}, \bibinfo{person}{Yasmine Badr},
  \bibinfo{person}{Jongsoo Park}, \bibinfo{person}{Jiyan Yang},
  \bibinfo{person}{Dheevatsa Mudigere}, {and} \bibinfo{person}{Ellie Wen}.}
  \bibinfo{year}{2022}\natexlab{}.
\newblock \bibinfo{title}{DHEN: A Deep and Hierarchical Ensemble Network for
  Large-Scale Click-Through Rate Prediction}.
\newblock
\showeprint[arxiv]{2203.11014}~[cs.IR]
\urldef\tempurl%
\url{https://arxiv.org/abs/2203.11014}
\showURL{%
\tempurl}


\bibitem[Zhang et~al\mbox{.}(2024)]%
        {Zhang_2024}
\bibfield{author}{\bibinfo{person}{Wei Zhang}, \bibinfo{person}{Dai Li},
  \bibinfo{person}{Chen Liang}, \bibinfo{person}{Fang Zhou},
  \bibinfo{person}{Zhongke Zhang}, \bibinfo{person}{Xuewei Wang},
  \bibinfo{person}{Ru Li}, \bibinfo{person}{Yi Zhou}, \bibinfo{person}{Yaning
  Huang}, \bibinfo{person}{Dong Liang}, \bibinfo{person}{Kai Wang},
  \bibinfo{person}{Zhangyuan Wang}, \bibinfo{person}{Zhengxing Chen},
  \bibinfo{person}{Fenggang Wu}, \bibinfo{person}{Minghai Chen},
  \bibinfo{person}{Huayu Li}, \bibinfo{person}{Yunnan Wu},
  \bibinfo{person}{Zhan Shu}, \bibinfo{person}{Mindi Yuan}, {and}
  \bibinfo{person}{Sri Reddy}.} \bibinfo{year}{2024}\natexlab{}.
\newblock \showarticletitle{Scaling User Modeling: Large-scale Online User
  Representations for Ads Personalization in Meta}. In
  \bibinfo{booktitle}{\emph{Companion Proceedings of the ACM Web Conference
  2024}} \emph{(\bibinfo{series}{WWW ’24})}. \bibinfo{publisher}{ACM},
  \bibinfo{pages}{47–55}.
\newblock
\href{https://doi.org/10.1145/3589335.3648301}{doi:\nolinkurl{10.1145/3589335.3648301}}


\bibitem[Zhuang et~al\mbox{.}(2025)]%
        {zhuang2025}
\bibfield{author}{\bibinfo{person}{Jinfeng Zhuang}, \bibinfo{person}{Yinrui
  Li}, \bibinfo{person}{Runze Su}, \bibinfo{person}{Ke Xu},
  \bibinfo{person}{Zhixuan Shao}, \bibinfo{person}{Kungang Li},
  \bibinfo{person}{Ling Leng}, \bibinfo{person}{Han Sun}, \bibinfo{person}{Meng
  Qi}, \bibinfo{person}{Yixiong Meng}, \bibinfo{person}{Yang Tang},
  \bibinfo{person}{Zhifang Liu}, \bibinfo{person}{Qifei Shen}, {and}
  \bibinfo{person}{Aayush Mudgal}.} \bibinfo{year}{2025}\natexlab{}.
\newblock \bibinfo{title}{On the Practice of Deep Hierarchical Ensemble Network
  for Ad Conversion Rate Prediction}.
\newblock
\showeprint[arxiv]{2504.08169}~[cs.LG]
\urldef\tempurl%
\url{https://arxiv.org/abs/2504.08169}
\showURL{%
\tempurl}


\end{thebibliography}

\clearpage
\clearpage
\appendix

\section{Upstream Model Architecture}
\subsection{User Tower}
The user tower is designed to ingest a wide range of user features, including sequences of user activities, user demographics, curated user interests, user counting features and pretrained representations, etc. 
The user tower output is a set of user embedding tokens with the same dimension. We further flatten them into one embedding followed by $\ell_2$ normalization.
\[
    \mathrm{embedding} = \mathrm{Norm}(\mathrm{Concat}(\mathrm{token}_{1}, \mathrm{token}_{2}, ..., \mathrm{token}_{n}) \label{user_embedding}
\]

\subsection{Pin Tower}
Similarly, the Pin tower ingests all ad Pin related features. This includes content metadata, aggregated counting features, pretrained representations and other relevant features that provide insights into ad content. The output dimension is set to be the same as user embeddings to facilitate the inner product computation in the self supervised learning stage. 

\subsection{Overall Interaction Tower}
The overall interaction tower captures interactions among tower outputs and other features such as context features and interaction features.

\section{Training and Inference}
\subsection{Dataset} We develop DERM embeddings from Pinterest's ad large-scale production datasets:
\begin{itemize}
    \item \textbf{Ad Engagement Dataset}:
    This training dataset is designed for click-through rate (CTR) prediction tasks in ad ranking, comprising ad events like impressions and clicks across various platforms, including Home Feed, RePin, and Search. 
    \item \textbf{Ad Conversion Dataset}: 
    This training dataset is intended for ads conversion rate (CVR) prediction tasks, comprising ad events such as impressions, clicks, and conversions. 
    \item \textbf{Features}: The feature set consists of features representing various aspects of user and item. They can be roughly categorized into aggregated counting features, categorical features, interaction features, sequences and embeddings. Of particular importance are the user sequence features, which include both onsite engagement sequences and offsite conversion sequences.
\end{itemize}
\subsection{Training} 
The upstream model is initially trained from scratch using data from an extended time window. Subsequently, on a daily basis, the model is incrementally trained with newly arrived data by loading the weights from the last model snapshot. This approach allows for the daily updating of entity representations.
\subsection{Inference} 
Each day, the model snapshot from the incremental training is used to generate entity embeddings from the same training data in an offline manner. To extend the availability of trained embeddings over more days, we can perform \textit{back inference} on the days within the batch training window using the latest model snapshot. While this approach may appear to introduce 'leakage,' it actually enriches the entity representations and has proven effective in enhancing downstream model performance. 
\section{Feature Processing and Serving}
Our feature processing pipeline addresses two key concerns with pre-trained entity representations: stability and coverage.

\subsection{Raw Feature Deduplication} The same user or ad may appear multiple times in a day, resulting in multiple embeddings. Analysis of their cosine similarities demonstrates minimal daily variations, so we retain the last embedding of the day.

\subsection{Feature Aggregation}
\label{sec:feature_aggregation}
To ensure coverage and stability, we merge new daily embeddings into the existing pool following aggregation heuristics. If a user or ad doesn't have a new embedding, we carry over the existing one.

\begin{itemize}
    \item Heuristic 1 - Accumulation: keep the latest embedding
    \item Heuristic 2 - Moving Average: weighted sum of the existing embedding and the newly inferred embedding
    
    $E_{\mathrm{agg}}(t) =wE_{\mathrm{agg}}(t-1) + (1 - w)E_{\mathrm{daily}}(t)$
    \item Heuristic 3 - Average Pooling: average of the latest two daily embeddings
\end{itemize}
We select the heuristics based on results from offline experiments. The moving average with $w=0.8$ yields the best result. More details are in Section \ref{sec:aggregation_heuristics_experiment}. 

We upload the aggregated embeddings to a key-value feature store for online serving.
To save cost, we keep embedding from users and ads that are active in the past three months. It achieves both good user and Pin coverage, and a high average day to day cosine similarity.

\subsection{Online Serving}
User embeddings are served through an online key-value feature store, while Pin embeddings are built into Pin indexing for ranking and retrieval.

\section{Downstream Model}

\subsection{Effects of Projection Layer}

Given that the time complexity of DCNv2 ~\cite{DCNv22021} is $O(d^2l)$, where $d$ is
the input dimension and $l$ is the number of layers, we introduce a projection layer
to reduce the dimensionality of the DERM embeddings. This step helps to reduce
infrastructure costs associated with the increased dimensions. However, it should be
noted that reducing the dimensionality may result in a performance drop.

We conducted experiments to evaluate the impact of the
projection layer on Pinterest's Related Pins (RP) Surface \footnote{An example page of Related Pins (RP) Surface: \url{https://www.pinterest.com/pin/68748667546/}. RP surface 
has the largest traffic among all other surfaces in Pinterest.}
model. A model with a projection layer of $512$ dimensions achieved the best tradeoff between
performance and infrastructure cost, with a $0.18\%$ ROC-AUC lift compared to
a model without the projection layer, which had a $0.19\%$ ROC-AUC lift.
Additionally, the model with the projection layer demonstrated a $5.68\%$ increase
in offline training throughput, leading to significant cost savings in online
inference infrastructure (approximately 212K US dollars in annual savings).
\section{Results in Offline Experiments in CVR Downstream Task }
We also conducted offline experiments to evaluate the impact of DERM
embeddings on the downstream CVR task.

\subsection{Number of DERM Embeddings}
\label{sec:number_of_derm_cvr}

Similarly, in this section,  we conducted experiments to evaluate the impact of different
DERM embeddings on the CVR task by testing four types of embeddings: user and item
embeddings from the upstream CTR model, and user and item embeddings from the
upstream CVR model. The results, presented in Table
\ref{tab:number_of_derm_embeddings_cvr}, demonstrate that the combination of user
and item embeddings from both models yields the best performance on Pinterest's CVR prediction.  
We can see that even with embeddings trained with the CVR dataset, there is significant offline AUC gain achieved for CVR task. The gain increased by 50\% combined with CTR embeddings, which comes from the cross domain transfer of CTR domain. 

\begin{table}[h]\centering
  \caption{Sensitivity of Number of DERM Embeddings on Downstream Task (CVR Prediction) evaluated by ROC-AUC and PR-AUC Lift. A 
  checkmark (\checkmark) indicates that the corresponding embedding is used in the model.
  An offline $0.05\%$ ROC-AUC lift is considered significant in Pinterest.}
  \label{tab:number_of_derm_embeddings_cvr}
  \scriptsize
  \begin{tabular}{c|c|c|c||c|c}\toprule
    \multicolumn{4}{c||}{Upstream Models} & \multicolumn{2}{c}{Downstream Task: CVR AUC } \\
    \midrule
    \multicolumn{2}{c|}{CTR} & \multicolumn{2}{c||}{CVR} & ROC-AUC Lift (\%) & PR-AUC Lift (\%) \\ \hline
    User & Item & User & Item &  \\ \hline
    \textbf{X} & \textbf{X} & \checkmark & \checkmark & 0.15 & 1.1 \\ \hline
    \checkmark & \checkmark & \checkmark & \textbf{X} & 0.17 & 1.79 \\ \hline
    \checkmark & \checkmark & \checkmark & \checkmark & 0.22 & 1.92 \\ \hline
    \bottomrule
  \end{tabular}
\end{table}

\end{document}